\documentclass[arxiv]{melba}
\usepackage{anyfontsize}
\usepackage{lmodern}

\usepackage{mwe} 

\usepackage{amsmath,amsfonts}
\usepackage{float}

\melbaid{YYYY: NNN}  
\doi{https://doi.org/10.59275/j.melba.2024-AAAA}
\melbaauthors{Deng and Wang et al}  
\volume{2}
\firstpageno{1337}  
\melbayear{2024}  
\datesubmitted{m1/yyyy}  
\datepublished{m2/2024}  

\melbaspecialissue{MEDICAL IMAGE COMPUTING AND COMPUTER ASSISTED INTERVENTION (MICCAI) 2024}
\melbaspecialissueeditors{Marleen de Bruijne, Tal Arbel, Ismail Ben Ayed, Hervé Lombaert}

\ShortHeadings{CTSpine1K: A Large-Scale Dataset for Spinal Vertebrae Segmentation in Computed Tomography}{Yang Deng and Ce Wang et al.}

\title{CTSpine1K: A Large-Scale Dataset for Spinal Vertebrae Segmentation  in Computed
Tomography}

\author{
	\firstname Yang Deng\aff{1,2,3},
         \firstname Ce Wang\aff{1,2},
         \firstname Yuan Hui\aff{1,2},
         \firstname Qian Li\aff{1,2},
         \firstname Jun Li\aff{1},
         \firstname Shiwei Luo\aff{4},
         \firstname Mengke Sun\aff{1},
         \firstname Quan Quan\aff{1},
         \firstname Shuxin Yang\aff{1},
         \firstname You Hao\aff{1,2},
         \firstname Pengbo Liu\aff{1},
         \firstname Honghu Xiao\aff{5},
         \firstname Chunpeng Zhao\aff{5},
         \firstname Xinbao Wu\aff{5},
         \firstname S. Kevin Zhou\aff{1,2,3}
}
\affiliations{
	\num 1 \addr Key Lab of Intelligent Information Processing of Chinese Academy of Sciences, Institute of Computing Technology, Chinese Academy of Sciences, Beijing, 100190, China \\
	\num 2 \addr Suzhou Institute of Intelligent Computing Technology, Chinese Academy of Sciences, Suzhou, 215028, China \\
	\num 3 \addr School of Biomedical Engineering, Division of Life Sciences and Medicine, University of Science and Technology of China, Hefei, Anhui, and also with the Center for Medical Imaging, Robotics, Analytic Computing \& Learning (MIRACLE), Suzhou Institute for Advanced Research, University of Science and Technology of China, Suzhou, Jiangsu, China\\
	\num 4 \addr Department of Radiology, Guangzhou First People's Hospital, School of Medicine, South China University of Technology, Guangzhou, 510006, China \\
	\num 5 \addr Beijing Jishuitan Hospital, Beijing, 100035, China \\
}

\abstract{
	Spine-related diseases have high morbidity and cause a huge burden of social cost. Spine imaging is an essential tool for noninvasively visualizing and assessing spinal pathology. Segmenting vertebrae in computed tomography (CT) images has always been the base of quantitative medical image analysis for clinical diagnosis and surgery planning of spine diseases. Current publicly
    available annotated datasets on spinal vertebrae are small in size. Due to the lack of a large-scale annotated spine image dataset, the mainstream deep learning-based segmentation methods, which are data-driven, are heavily restricted. In this paper, we introduce a large-scale spine CT dataset called CTSpine1K, curated from multiple sources for vertebra segmentation, which contains 1,005 CT volumes with over 500,000 labeled vertebrae slices and 11,172 vertebrae belonging to different spinal conditions. Based on this dataset, we conducted several spinal vertebrae segmentation experiments to set the first benchmark. We believe that this large-scale dataset will facilitate further research in many spine-related image analysis tasks, including but not limited to vertebrae segmentation, labeling, 3D spine reconstruction from biplanar radiographs, and image superresolution and enhancement.
	Our dataset are publically available at~\url{https://xnat.bmia.nl/data/archive/projects/africai_miccai2024_ctspine1k} and ~\url{https://github.com/MIRACLE-Center/CTSpine1K}.}

\keywords{Spine Dataset, Vertebrae Segmentation,  Computed Tomography (CT), Medical Imaging}

\begin{document}

\twocolumn[\maketitle]

\section{Introduction}
\enluminure{T}{he} \emph{spine} is an important part of the musculoskeletal system, sustaining body mobility and protecting the spinal cord, the most important neural pathway in the body \citep{cai2020computational, sekuboyina2020verse}. From top to bottom along the trunk of the human body, the spine consists of 7 cervical vertebrae (C1-C7), 12 thoracic vertebrae (T1-T12), 5 lumbar vertebrae (L1-L5), 1 sacral vertebra, and 1 caudal vertebra. Note that people are containing L6 (resulting from sacral lumbarization) or losing L5 (resulting from lumbar sacralization) with a rare occurrence in a population.

\begin{figure*}[ht]
        \begin{center}
            \includegraphics[width=0.95\linewidth]{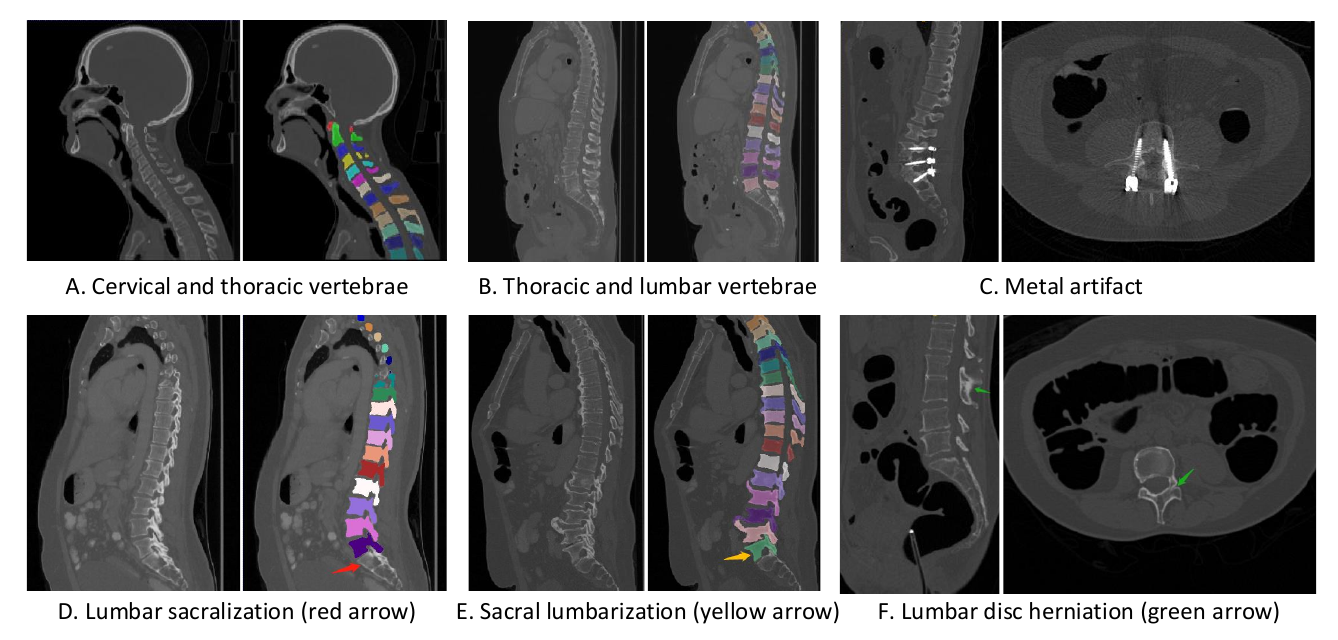}
        \end{center}
        \caption{Spine CT image examples with various conditions.}
        \label{fig:network}
\end{figure*}

Each vertebra is at risk of disease due to bearing the load of the human body~\citep{faruqi2018vertebral}. Spine-related diseases, including degenerative changes, spinal inflammation, spinal tumors, spinal tuberculosis, and spinal infection, have a high morbidity and cause a huge burden of social cost. In clinical practice, common spinal diseases include mainly degenerative diseases of the spine and disc herniation~\citep{baleriaux2004spinal}. Degenerative diseases of the spine common in the elderly include cervical spondylosis, lumbar spondylosis, and thoracic spinal stenosis. Disc herniation includes cervical disc herniation that involves the cervical spine, thoracic disc herniation that involves the thoracic spine, and common lumbar disc herniation, among which lumbar disc herniation is the most common. Lumbar degenerative diseases are also the most common in the clinic, with lumbar spinal stenosis, lumbar spondylolisthesis, and lumbar instability. Treatment varies with the disease entity, and the clinical scenario can be nonspecific~\citep{li2015spinal}.

Spinal imaging via computed tomography (CT), magnetic resonance imaging (MRI), radiography, ultrasound, positron emission tomography (PET), and other radiologic imaging modalities is essential for noninvasively visualizing and assessing spinal pathology. Computational methods support and enhance a physician’s ability to utilize these imaging techniques for diagnosis, noninvasive treatment, and intervention in clinical practice. Analysis algorithms developed in the field of computer vision, computer graphics, signal processing, and machine learning have been adapted to analyze spinal images~\citep{li2015spinal}. Conventionally, CT is preferred to study the spine due to a high bone-soft tissue contrast. There are diverse image appearance variations due to differences in vertebral position, metal artifacts and spinal diseases, etc., challenging the analysis algorithms. Fig.~\ref{fig:network} gives some examples of these various conditions.

Spinal or vertebral image segmentation in CT is crucial in all applications regarding automated quantification of spinal morphology and pathology. Over the recent years, deep learning has achieved remarkable success in various medical imaging applications \citep{zhou2021review,zhou2019handbook} and many automated spine image segmentation approaches have been proposed \citep{lessmann2019iterative,payer2020coarse}. However, all these approaches are data-dependant and have either been validated on private datasets or small public datasets. Considering SpineWeb~\footnote[1]{http://spineweb.digitalimaginggroup.ca/}, a popular archive for multi-modal spine data, it lists only two CT datasets: CSI2014 \citep{yao2012detection} and xVertSeg \citep{korez2015framework}, both of which only contain dozens of CT scans. Therefore, those approaches are heavily restricted. To address the concerns of large-scale data availability, Sekuboyina et al. \citep{sekuboyina2020verse} organized the Large Scale \underline{Ver}tebrae \underline{Se}gmentation Benchmark(VerSe) as a challenge in conjunction with the International Conference on Medical Image Computing and Computed Assisted Intervention (MICCAI) 2019 and MICCAI 2020. With VerSe'19~\footnote[2]{https://verse2019.grand-challenge.org/}, they released into the public domain a diverse dataset of 160 spine multi-detector CT scans with 1,735 vertebrae (120 seen CT scans and 40 hidden CT scans). For VerSe'20~\footnote[3]{https://verse2020.grand-challenge.org/}, the upgraded version of VerSe'19, they released 300 CT scans, the largest public spine CT dataset to date.  These two datasets provide the ground truth of each vertebra and are currently the most used dataset for vertebrae segmentation. Nonetheless, these datasets are still small. Further, most CT scans from VerSe'19 and VerSe'20 datasets are `cropped', only containing a small area of the spine and abandoning other information about surrounding organs.

To advance the research in spinal image analysis, we hereby present a large-scale and comprehensive dataset: CTSpine1K.  We collect and annotate a large-scale spinal vertebrae CT dataset from multiple domains and different manufacturers, totaling 1,005 CT volumes (over 500,000 labeled slices and over 11,000 vertebrae) of diverse appearance variations. We carefully design an exquisite and unified annotation pipeline to ensure the quality of annotations. To the best of our knowledge, our CTSpine1K dataset is the largest publicly available annotated spine CT dataset. We evaluate the dataset's quality by conducting benchmark experiments for vertebrae segmentation.

\begin{figure*}[htbp]
		\begin{center}
			\includegraphics[width=0.95\linewidth]{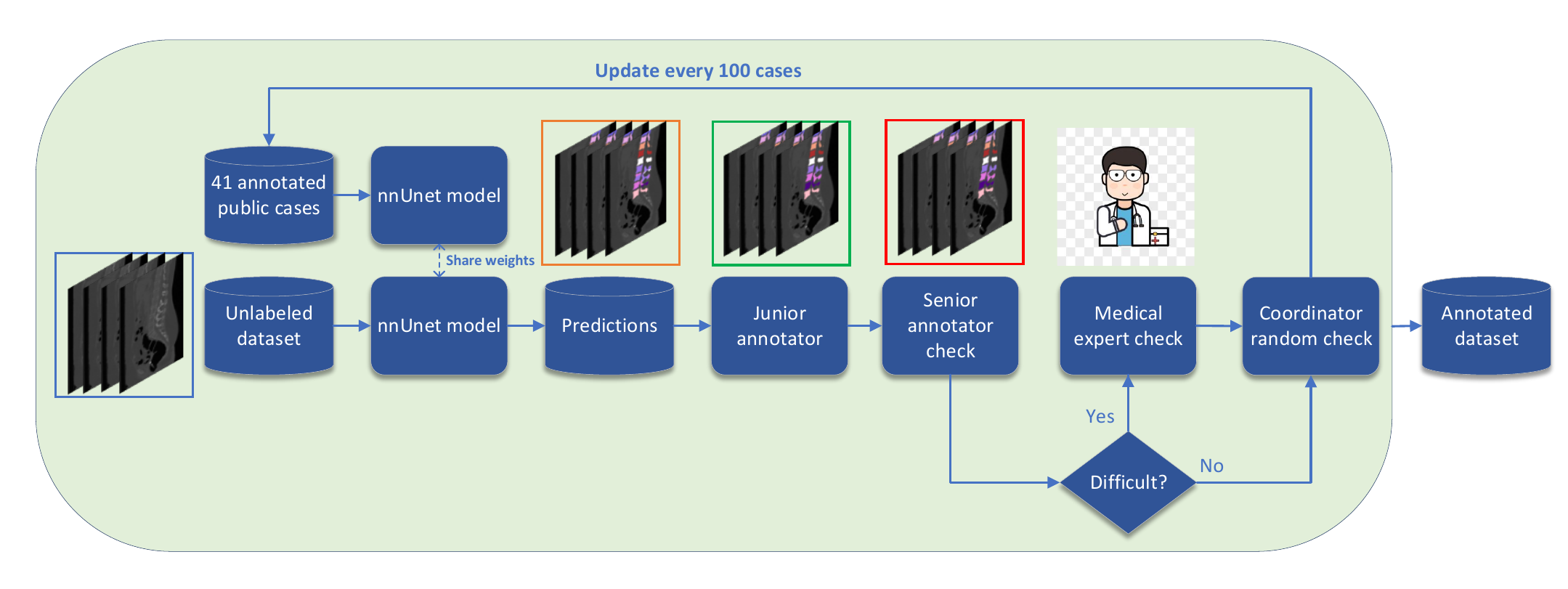}
		\end{center}
		\caption{The proposed annotation pipeline.}
		\label{framework}
\end{figure*}
    


\section{The CTSpine1K Dataset}
\subsection{Data Collection}
To build a comprehensive spine image dataset that replicates practical appearance variations, we curate a large-scale CT dataset of spinal vertebrae from the following four open sources.

\underline{COLONOG.} This sub-dataset comes from
COLONOGRAPHY dataset related to a CT colonography trial \citep{johnson2008accuracy}. We randomly select one of the two positions (we will open source the codes used for selection), which have similar information for each patient in our dataset. There are 825 CT scans, and they are in a Digital Imaging and Communication in Medicine (DICOM) format.

\underline{HNSCC-3DCT-RT.} This sub-dataset contains three-dimensional (3D) high-resolution fan-beam CT scans collected during pre-, mid-, and post-treatment using a Siemens 16-slice CT scanner with a standard clinical protocol for 31 head-and-neck squamous cell carcinoma (HNSCC) patients \citep{bejarano9head}. These images are in a DICOM format.

\underline{MSD T10.} This sub-dataset comes from the 10th Medical Segmentation Decathlon \cite{simpson2019large}. To attain more slices containing the spine, we select task03\_liver dataset consisting of 201 cases. These images are in a Neuroimaging Informatics Technology Initiative (NIfTI) format (https://nifti.nimh.nih.gov/nifti-1).

\underline{COVID-19.} This sub-dataset consists of non-enhanced chest CTs from 632 patients with COVID-19 infections. The images were acquired at the point of care in an outbreak setting from patients with Reverse Transcription Polymerase Chain Reaction (RT-PCR) confirmation for the presence of SARS-CoV-2 \citep{harmon2020artificial}. We pick 40 scans with the images stored in a NIfTI format.

We reformat all DICOM images to NIfTI to simplify data processing and de-identify images, meeting contributing sites' institutional review board (IRB) policies. More details for those sub-datasets can be found in \citep{johnson2008accuracy, bejarano9head,simpson2019large,harmon2020artificial}. All existing sub-datasets are under the Creative Commons Attribution-NonCommercial-ShareAlike 4.0 International License (CC BY-NC-SA 4.0); we will keep the license unchanged. It should be noted that for sub-dataset task03\_liver and sub-dataset COVID-19, we only choose a part of cases from them, and in all these data sources, we exclude those cases of very low quality. The overview of our dataset and the thorough comparison with the VerSe Challenge dataset can be seen in Table~\ref{tab1}.

\begin{table*}[ht]
    \centering
    \caption{Overview of our large-scale CTSpine1K dataset. Ticks[\checkmark] in the table refer to the fact that this dataset contains some special cases with sacral lumbarization or lumbar sacralization, and we list their IDs in the open source. 'Tr$\backslash$Ts\_pu$\backslash$Ts\_pr' donates training$\backslash$test\_public$\backslash$test\_private. We exclude the metal-artifact cases due to the difficulty of labeling them.}\label{tab1}
    \resizebox{\linewidth}{!}{
    \begin{tabular}{l|c|cccc|c}

    \toprule[1pt].
    Dataset name &VerSe'19,20 [\checkmark] & COLONOG [\checkmark] & HNSCC-3DCT-RT&  MSD T10 [\checkmark] &COVID-19 &CTSpine1K [\checkmark]\\
    \hline
    \hline
     \# of patients(n) & 41 & 784& 31 & 150 & 40 & 1005 \\
    \# of labeled vertebrae(n)& 511  &8022 & 443 & 2105&602 &11172\\
    \# of Tr$\backslash$Ts\_pu$\backslash$Ts\_pr(n)&- &480$\backslash$152$\backslash$152  &20$\backslash$5$\backslash$6 &90$\backslash$30$\backslash$30&20$\backslash$10$\backslash$10  & 610$\backslash$197$\backslash$198\\
    \# of cervical vertebrae(n)& 17  &4 & 217 & 132&66 &419\\
    \# of thoracic vertebrae(n)& 285  &4039 & 226 &1198 &479 &5942\\
    \# of lumbar vertebrae(n)& 194  & 3911& 0 &744 & 57&4712\\
    \# of sacral lumbarization&  8 & 16& 0 &1 &0 &17\\
    \# of lumbar sacralization&  3 & 9& 0 &2 &0 &11\\
    Mean spacing(mm) & (0.81$\pm$.1, 0.81$\pm$.1, 0.88$\pm$.16)&(0.75$\pm$.08, 0.75$\pm$.08, 0.81$\pm$.16)&(1.09$\pm$.14, 1.09$\pm$.14, 2.0)&(0.77$\pm$.1, 0.77$\pm$.1, 1.55$\pm$1.22)&(0.79$\pm$.09, 0.79$\pm$.09, 4.5$\pm$1.34) &(0.76$\pm$.11, 0.76$\pm$.11, 1.10$\pm$.94)\\
    Mean size &(512, 512, 640$\pm$197)&(512, 512, 542$\pm$58) &(512, 512, 202$\pm$19)&  (512, 512, 478$\pm$264)&(512, 512, 93$\pm$76)& (512,512,504$\pm$155)\\
     Age& 56.2 $\pm$ 17.6 &56.7 $\pm$ 6.1 years&64.3 $\pm$ 12.8 years&Unknown &Unknown&-\\
     Manufacturer&Siemens,GE,Philips,Toshiba &Siemens,GE,Philips,Toshiba &Siemens&Unknown&Unknown & -\\
    Source and Year & URL\footnotemark[1]$^,$\footnotemark[2] 2019,2020 &\citep{johnson2008accuracy}  &\citep{bejarano9head} &\citep{simpson2019large} & \citep{harmon2020artificial}  &- \\
    \toprule[1pt]

    \end{tabular}
    }
    \end{table*}
    \footnotetext[1]{https://verse2019.grand-challenge.org/}
    \footnotetext[2]{https://verse2020.grand-challenge.org/}

\subsection{Data Annotation}
We design a unified and rigorous labeling standard and pipeline before annotation because medical image annotation is a highly time-consuming and subjective task. The annotation pipeline is shown in Fig.~\ref{framework}.

To reduce the annotation workload, we first use the public dataset from VerSe'19 and VerSe'20 Challenges to train a segmentation network using the nnUnet algorithm \citep{isensee2021nnu}. As mentioned earlier, most of the VerSe Challenge samples are 'cropped', abandoning the information about surrounding structures such as organs. Therefore, we selected cases with complete CT images (without cropping) and the same image spacing between images and their corresponding ground truth, totaling 41 cases that could be used. Then, for an image to be annotated, we invoke the trained segmentation model to predict segmentation masks and invite some junior annotators to refine the labels based on the prediction results. All these refined labels by junior annotators are checked by two senior annotators for further refinement. If the senior annotator finds it challenging to determine the annotations, these data will be sent to one of the trained spine surgeons, whose image-reading experience averages 12 years. Finally, all of these annotated labels undergo a random double-check by coordinators to ensure the final quality of annotations.

If there exist any wrong cases in double-check, they are corrected by annotators. The human-corrected annotations and their corresponding images are then added to the training data to retrain a more powerful model. To speed up the annotation process, we update the database every 100 cases and retrain the deep learning model. The process is iterative until the annotation task is finished. The whole annotation process is operated with ITK-SNAP \citep{py06nimg} software. Segmentation masks are also saved in a NIfTI format. In total, we have annotations for 1005 CT volumes.






\subsection{Usage Notes}
The data within this work is licensed under the CC BY-NC-SA 4.0 International License. The data can be freely downloaded and used for your own research purposes, but we kindly ask investigators to cite this paper in their publications. The data are suitable for visualization in a variety of software, including 3D Slicer \citep{yushkevich2006user} and ITK-SNAP \citep{py06nimg}.

\subsection{Code availability}
We provide the Python scripts and our trained model on GitHub~\footnote[1]{https://github.com/MIRACLE-Center/CTSpine1K}, which can serve as a starting point for the community to march on future development based on our CTSpine1K dataset. All the datasets were uploaded to the XNAT~\footnote[2]{https://xnat.bmia.nl/data/archive/projects/africai\_miccai2024\_ctspine1k} platform following MICCAI 2024 AFRICAI Imaging Repository White Paper \citep{starmans_2024_10816769}.

\section{Benchmarking Experiment}
Based on CTSpine1K, we use a fully supervised method to train a deep network for spinal vertebrae segmentation to establish a benchmark. In recent years, the nnUnet model has achieved better results than other methods in many medical image segmentation tasks and has become the acknowledged baseline in medical image segmentation \cite{isensee2021nnu}. nnUnet is essentially a U-Net \cite{ronneberger2015u}, but with specific network architecture, design parameters, and training parameters self-adapted to the dataset's characteristics, together with powerful data augmentation. Therefore, we choose nnUnet as the benchmarking model for vertebrae segmentation. Due to our dataset's huge amount of high-resolution 3D images, we use the 3D full-resolution U-net architecture. More details for the nnUnet model can be seen in \cite{isensee2021nnu}.

\begin{table*}[htbp]
    \centering
    \footnotesize
\setlength{\abovecaptionskip}{0cm}
\setlength{\belowcaptionskip}{-0.1cm}
    \caption{The DSC and HD95 (in mm) results. The Mean of Vertebrae indicates the average of each vertebra (multi-class), and the Mean of Spine indicates the average of the whole spine (binary classification). }
    \begin{center}
            \resizebox{\linewidth}{!}{
            \begin{tabular}{c|c|c|c|c|c|c|cc}
                \toprule[1pt]
                \multirow{2}*{Label}&Test\_public&Test\_private&Test\_VerSe &\multirow{2}*{Label}&Test\_public&Test\_private&Test\_VerSe \\
                \cline{2-4}\cline{6-8}&DSC/HD95&DSC/HD95& DSC/HD95 & &DSC/HD95& DSC/HD95&DSC/HD95\\
                \hline
                \hline
                C1&.951/1.55&.957/1.34&.446/1.69&T7&.881/4.43&.764/5.98&.400/8.67\\
                C2&.974/1.15&.975/1.22&.923/1.20&T8&.856/6.96&.772/6.23&.395/8.74\\
                C3&.934/8.73&.945/7.55&.605/1.32&T9&.949/1.76&.925/3.18&.597/6.33\\
                C4&.946/1.90&.608/3.65&.613/0.95&T10&.952/1.59&.951/1.85&.813/3.83\\
                C5&.573/2.43&.726/3.13&.231/0.95&T11&.959/1.49&.961/1.81&.887/3.36\\
                C6&.802/4.51&.824/2.38&.121/1.07&T12&.931/1.49&.965/1.73&.889/1.00\\
                C7&.930/2.18&.899/2.16&.228/4.04&L1&.965/1.27&.883/1.76&.925/2.34\\
                T1&.952/1.60&.927/2.36&.623/2.53&L2&.965/1.50&.913/1.91&.902/4.83\\
                T2&.960/1.43&.932/2.59&.883/2.36&L3&.963/1.61&.952/2.05&.830/7.17\\
                T3&.955/1.64&.798/7.47&.817/4.14&L4&.965/2.32&.945/2.61&.795/8.94\\
                T4&.899/3.83&.831/3.64&.723/6.44&L5&.969/1.49&.943/2.45&.723/10.31\\
                T5&.681/6.67&.830/4.90&.614/9.05&L6&0/66.21&0/80.33&0/45.16\\
                T6&.814/5.20&.762/5.99&.483/9.21&&&\\
                \hline
                \hline
                Mean of Vertebrae&.869/5.40&.840/6.41&.619/6.23&Mean of Spine&.985/1&.984/1&.929/3.08\\
                \hline
            \end{tabular}
            }
    \end{center}
    \label{tab2}
\end{table*}

\begin{figure*}[ht]
\begin{minipage}[t]{0.16\textwidth}
\centering
\includegraphics[width=\textwidth]{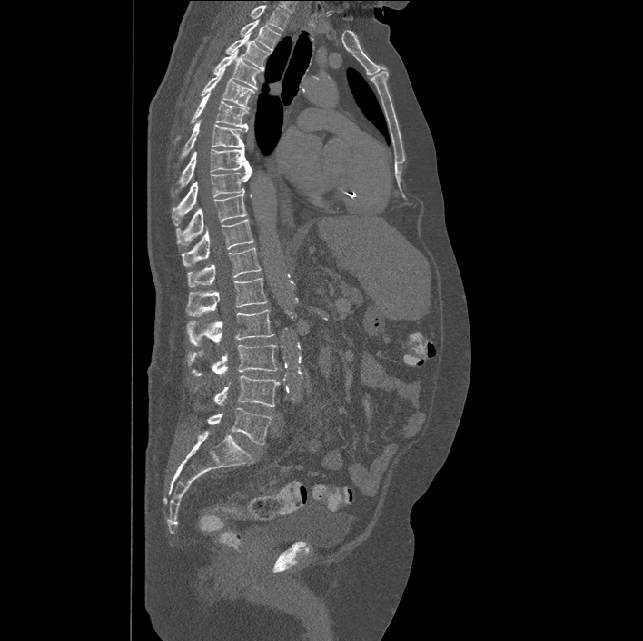}
VerSe'
\end{minipage}
\begin{minipage}[t]{0.16\textwidth}
\centering
\includegraphics[width=\textwidth]{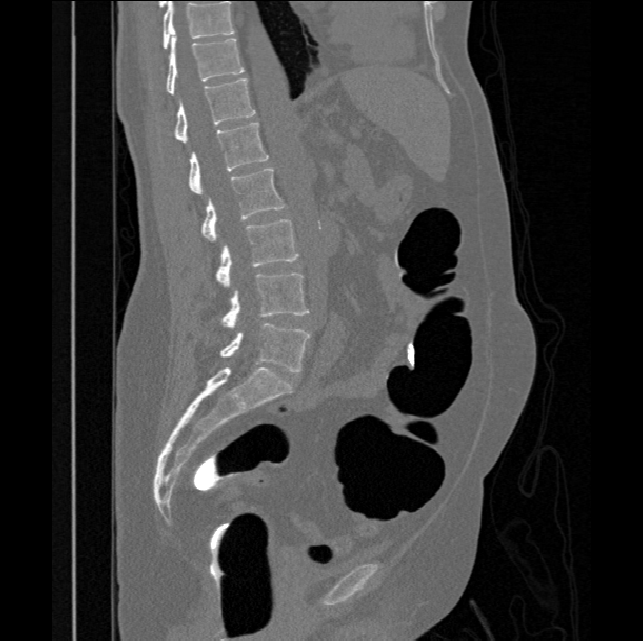}
COLONOG
\end{minipage}
\begin{minipage}[t]{0.16\textwidth}
\centering
\includegraphics[width=\textwidth]{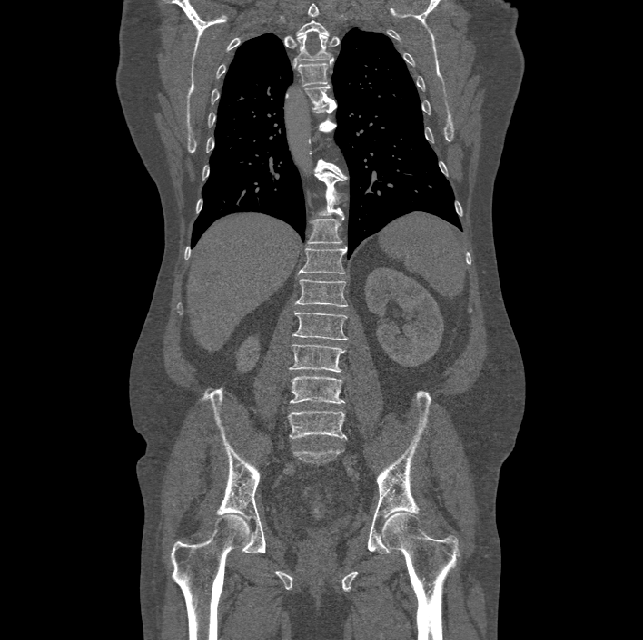}
VerSe'
\end{minipage}
\begin{minipage}[t]{0.16\textwidth}
\centering
\includegraphics[width=\textwidth]{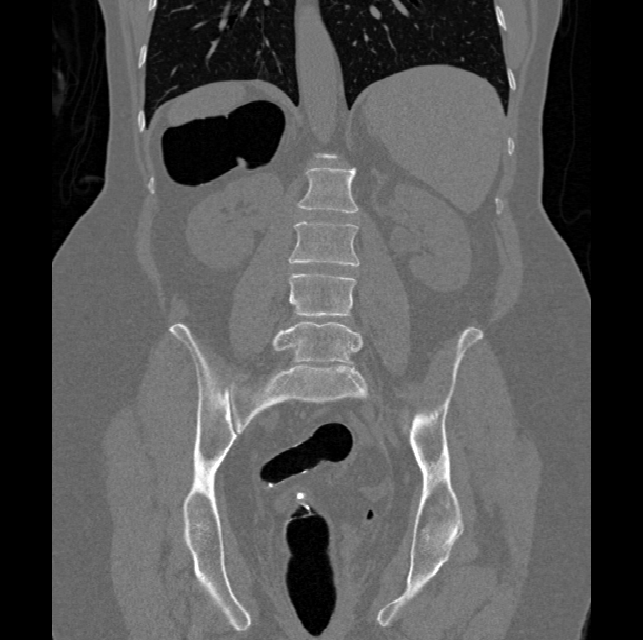}
COLONOG
\end{minipage}
\begin{minipage}[t]{0.16\textwidth}
\centering
\includegraphics[width=\textwidth]{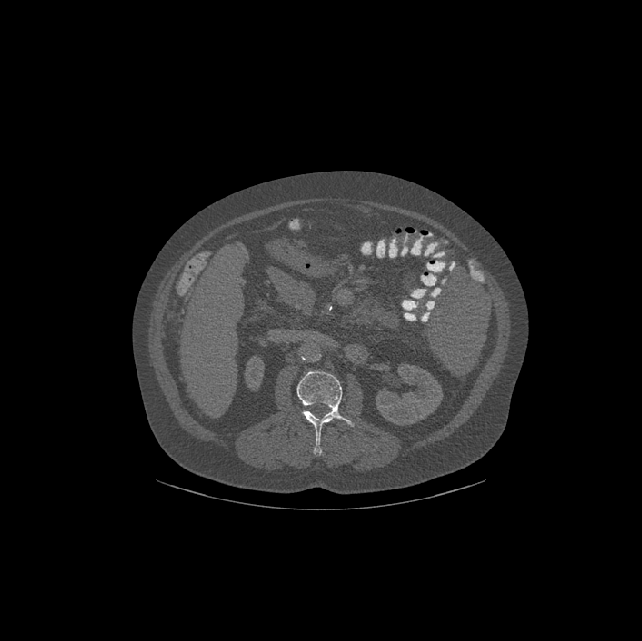}
VerSe'
\end{minipage}
\begin{minipage}[t]{0.16\textwidth}
\centering
\includegraphics[width=\textwidth]{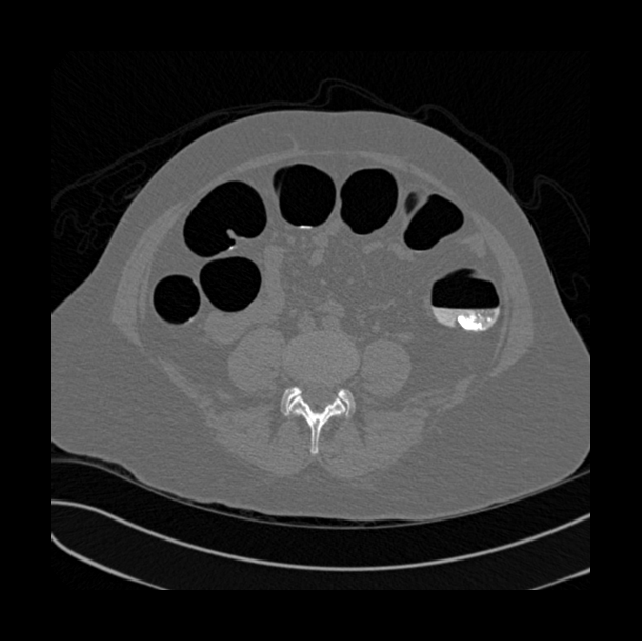}
COLONOG
\end{minipage}
\caption{The difference between the COLONGO dataset and VerSe'dataset. }
\label{difference}
\end{figure*}

\subsection{Experimental Setup}
\noindent \textbf{Data Split.} Our dataset contains 1005 3D CT volumes (on average, each scan has 504 slices and 11 labeled vertebrae) with over 500,000 labeled vertebrae slices (size of 512x512). The CTSpine1K dataset is separated into a training dataset (610 subjects), a public test dataset (197 subjects), and a private test dataset (198 subjects). More details about the data split can be seen in Table~\ref{tab1}. Aiming to compare the domain differences between our annotated dataset and the VerSe Challenge datasets, we use the 41 public cases from the VerSe Challenges as Test\_VerSe.

\noindent \textbf{Evaluation Metrics.} Every vertebra (from C1 to L6) is labeled with integer values from 1 to 25. Here, we use two ubiquitous metrics prevalent in the medical image segmentation domain: (i) Dice Coefficient (DSC). DSC is computed per label:  $2|A*B|/(|A| + |B|)$, where A is the set of foreground voxels of a certain label, and B is the predicted mask.
(ii) 95th percentile Hausdorff distance (HD95) in mm. HD measures the local maximum distance between the two surfaces constructed from the ground truth and predicted segmentation map at the 95th percentile.

\noindent \textbf{Implementation Details.}
We train the nnUnet model for 1,000 epochs, keeping the training configuration, such as the learning rate and data augmentation, etc., the same as the original settings in reference \citep{isensee2021nnu}. Due to the limit of computing sources, we use \textit{all} folds rather than run five-fold cross-validation while training. The experiments are implemented using Pytorch on an RTX 3090 GPU. The training time for the last round is around 15 days.

\begin{figure*}[htbp]
    \begin{minipage}[t]{0.16\textwidth}
    \centering
    \includegraphics[width=\textwidth]{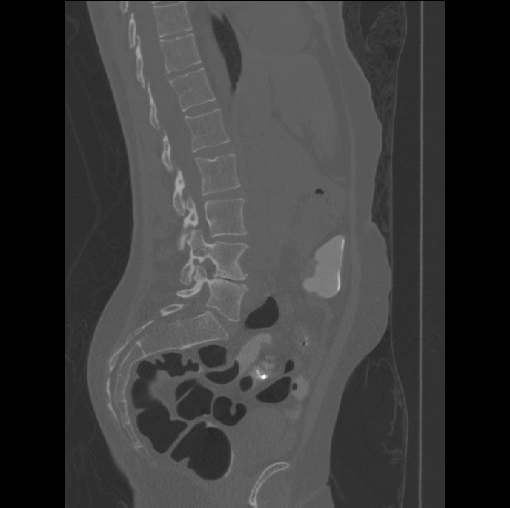}
    \includegraphics[width=\textwidth]{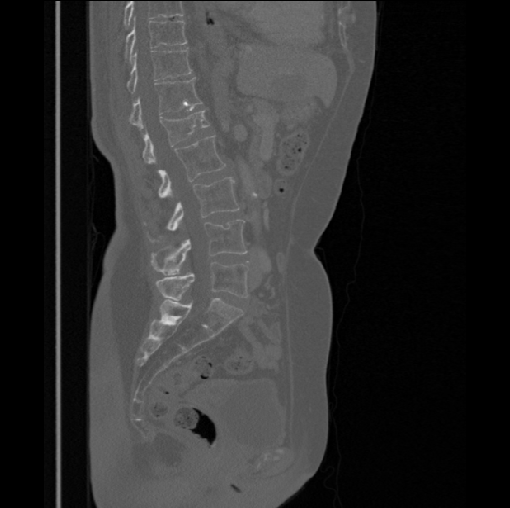}
    \includegraphics[width=\textwidth]{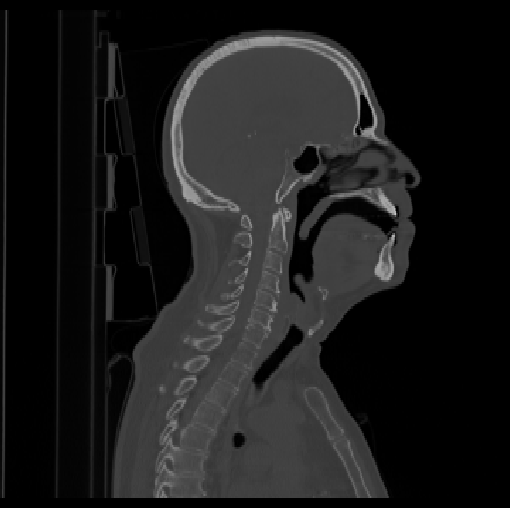}
    \includegraphics[width=\textwidth]{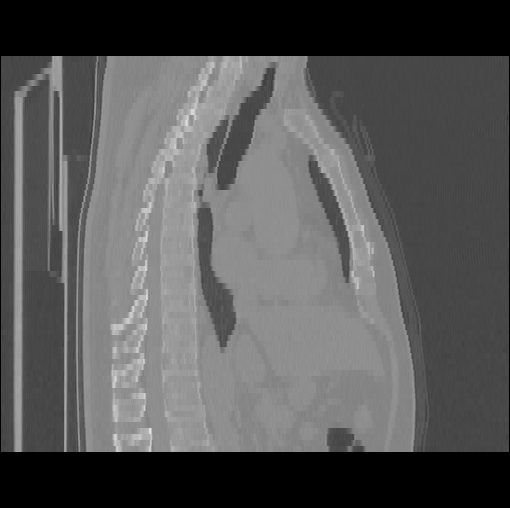}
    \includegraphics[width=\textwidth]{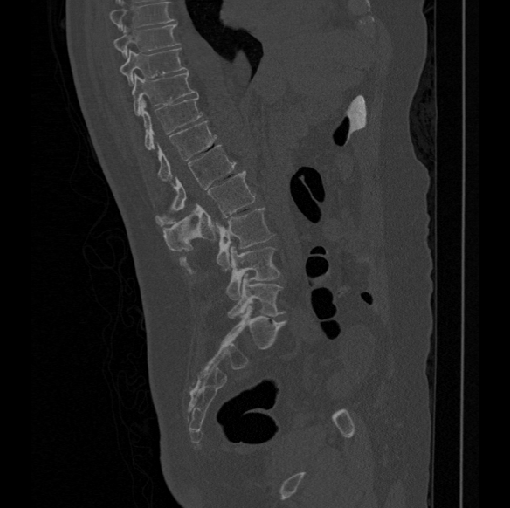}
    Images
    \end{minipage}
    \begin{minipage}[t]{0.16\textwidth}
    \centering
    \includegraphics[width=\textwidth]{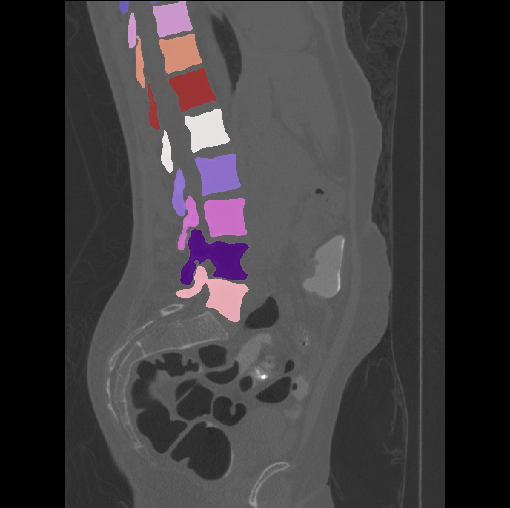}
    \includegraphics[width=\textwidth]{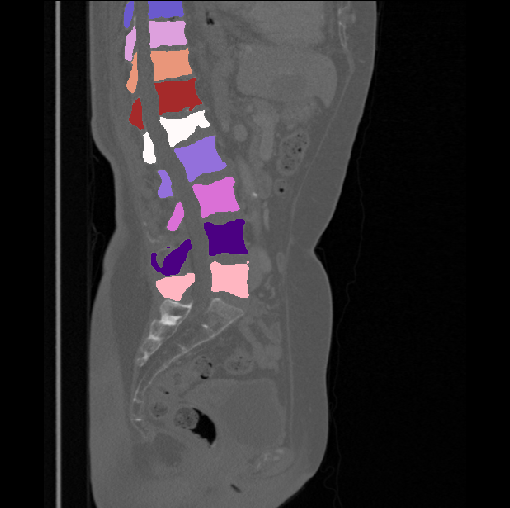}
    \includegraphics[width=\textwidth]{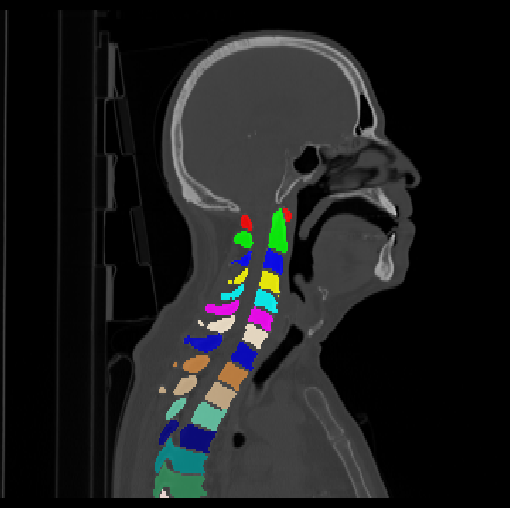}
    \includegraphics[width=\textwidth]{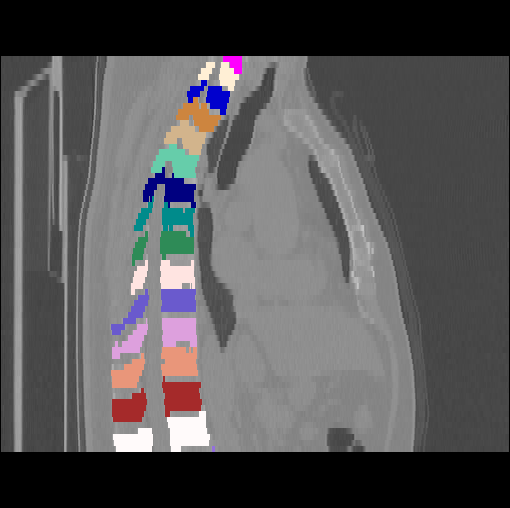}
    \includegraphics[width=\textwidth]{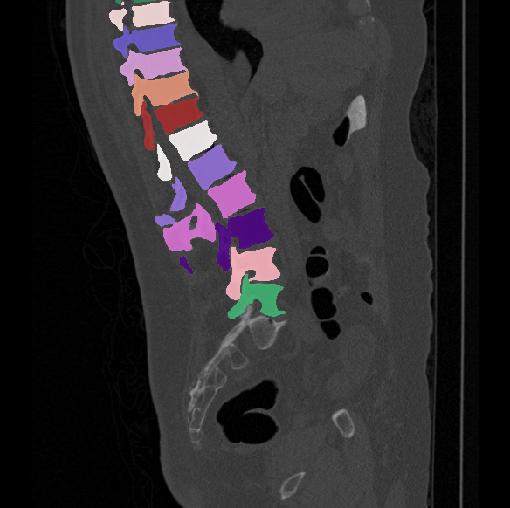}
    GT
    \end{minipage}
    \begin{minipage}[t]{0.16\textwidth}
    \centering
    \includegraphics[width=\textwidth]{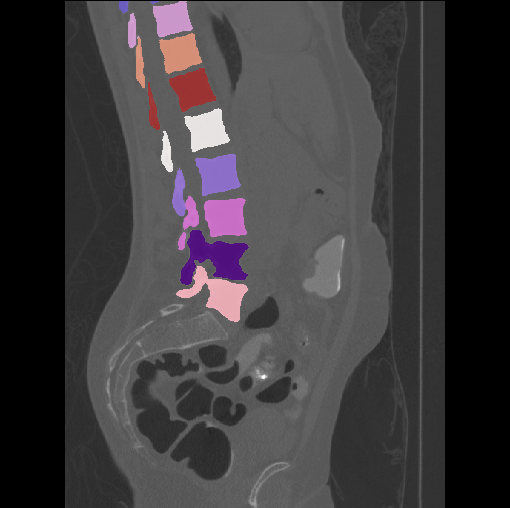}
    \includegraphics[width=\textwidth]{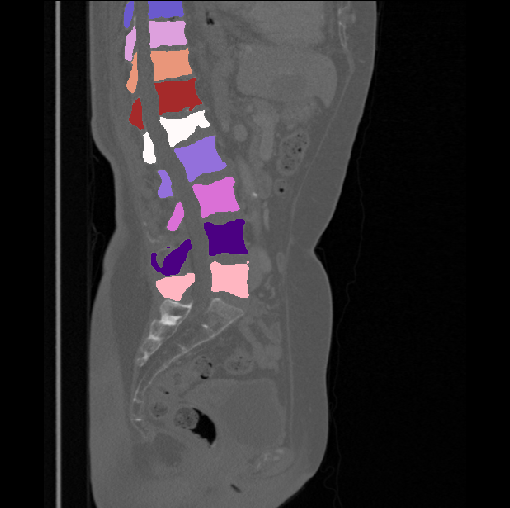}
    \includegraphics[width=\textwidth]{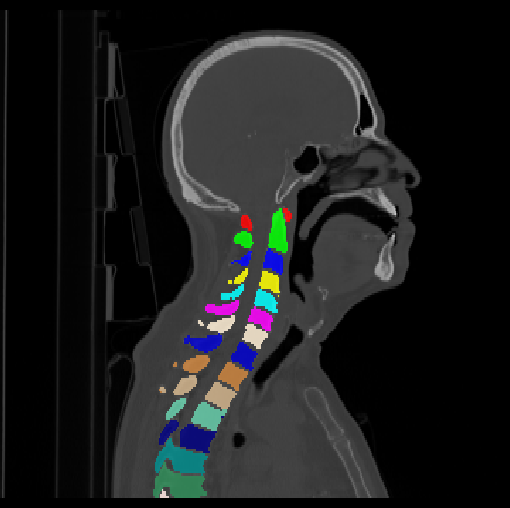}
    \includegraphics[width=\textwidth]{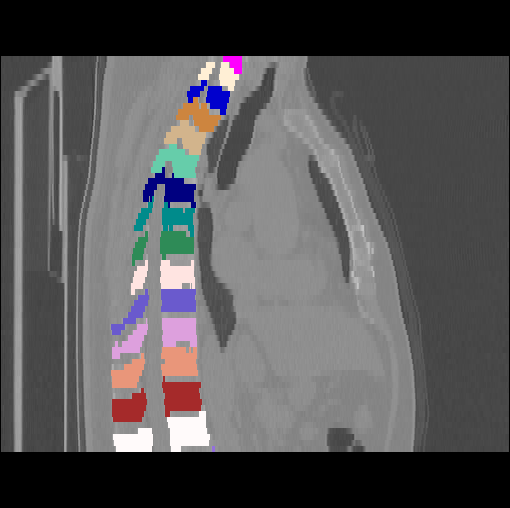}
    \includegraphics[width=\textwidth]{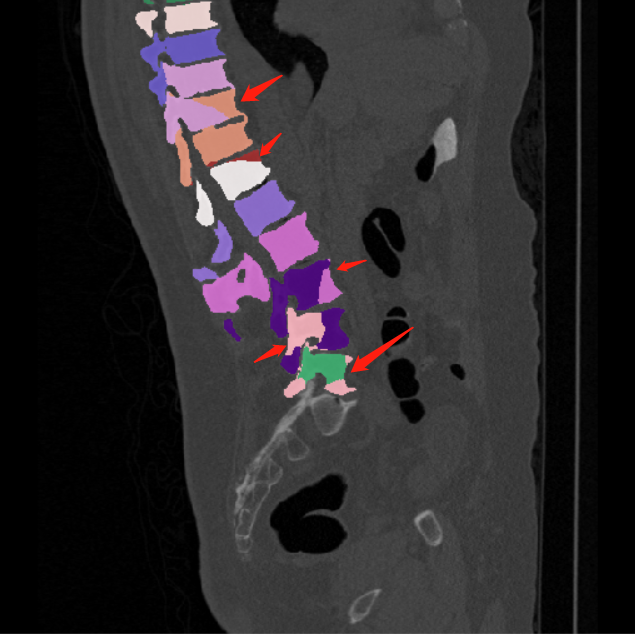}
    Predictions
    \end{minipage}
    \begin{minipage}[t]{0.16\textwidth}
    \centering
    \includegraphics[width=\textwidth]{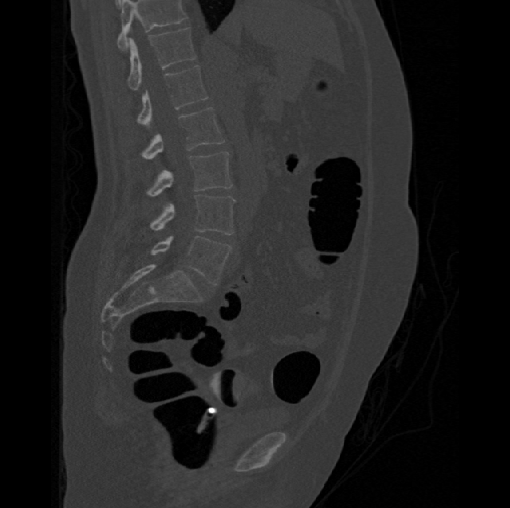}
    \includegraphics[width=\textwidth]{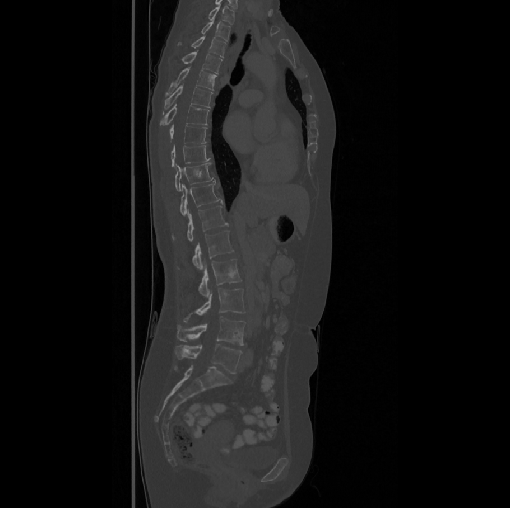}
    \includegraphics[width=\textwidth]{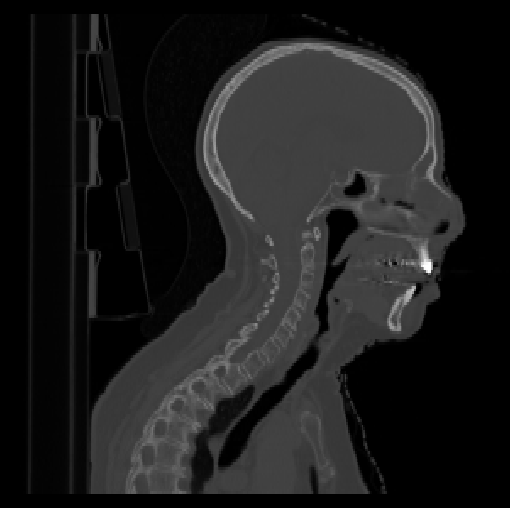}
    \includegraphics[width=\textwidth]{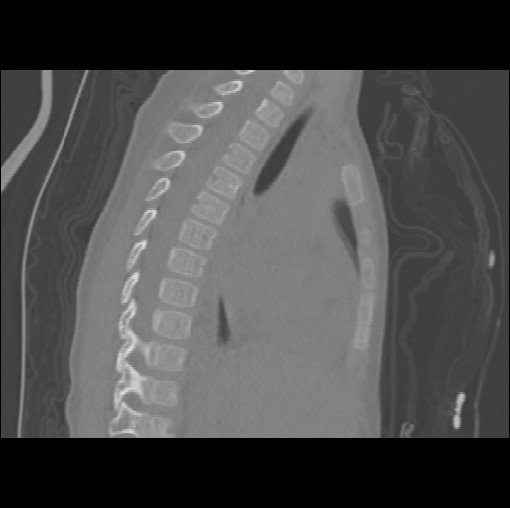}
    \includegraphics[width=\textwidth]{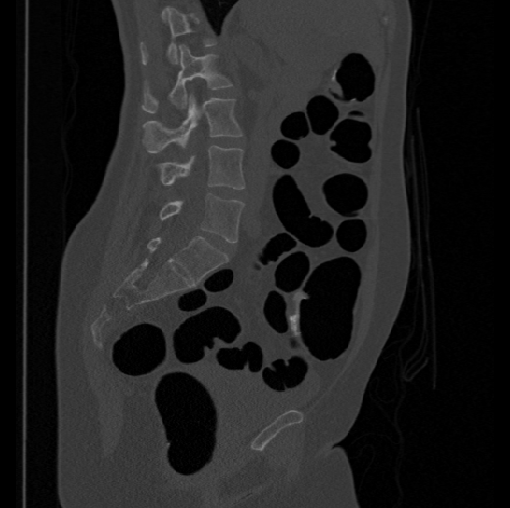}
    Images
    \end{minipage}
    \begin{minipage}[t]{0.16\textwidth}
    \centering
    \includegraphics[width=\textwidth]{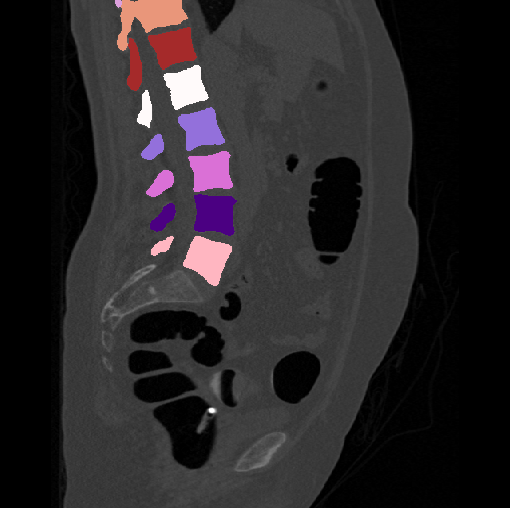}
    \includegraphics[width=\textwidth]{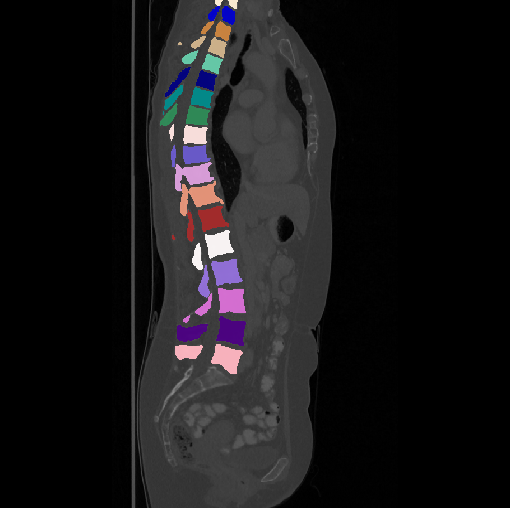}
    \includegraphics[width=\textwidth]{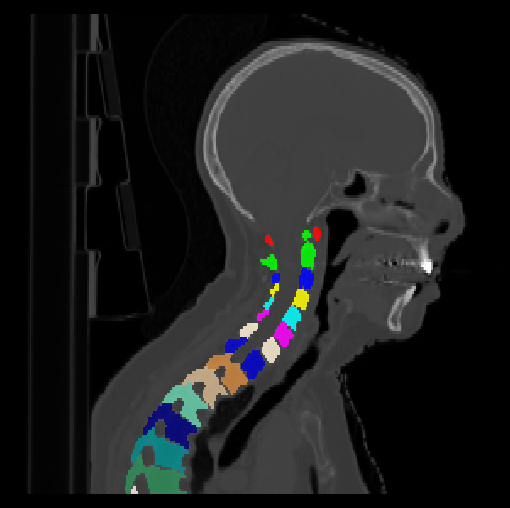}
    \includegraphics[width=\textwidth]{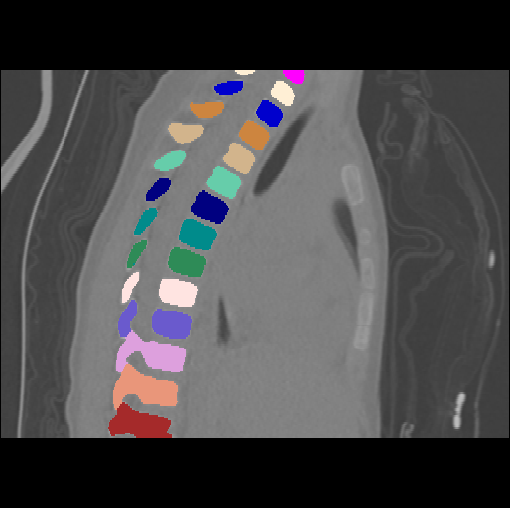}
    \includegraphics[width=\textwidth]{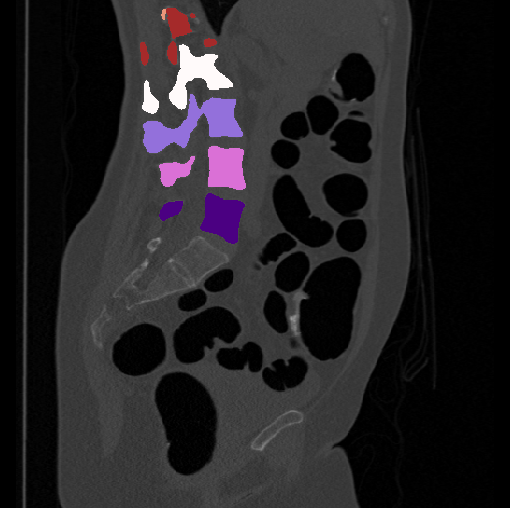}
    GT
    \end{minipage}
    \begin{minipage}[t]{0.16\textwidth}
    \centering
    \includegraphics[width=\textwidth]{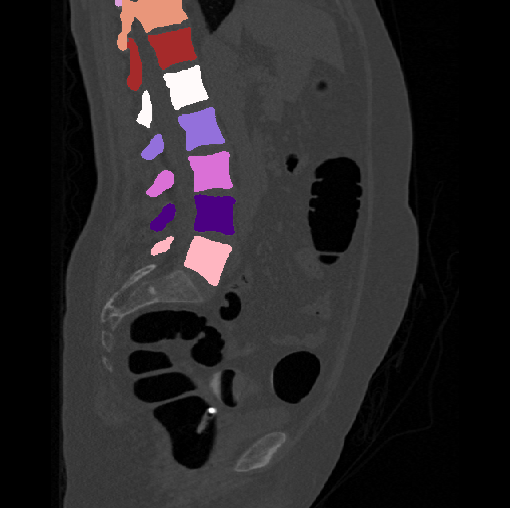}
    \includegraphics[width=\textwidth]{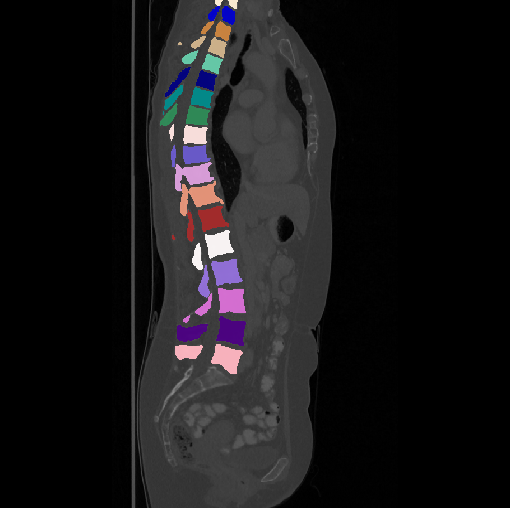}
    \includegraphics[width=\textwidth]{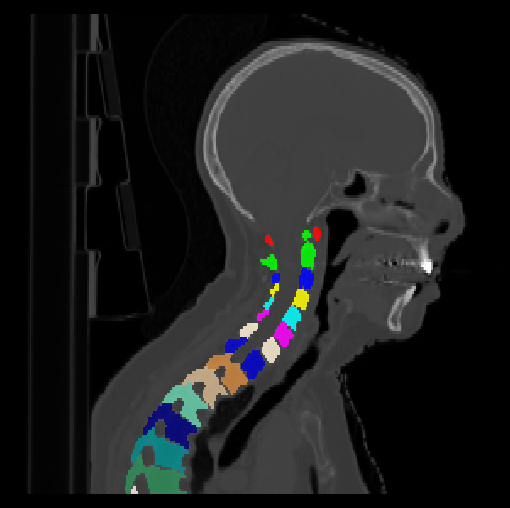}
    \includegraphics[width=\textwidth]{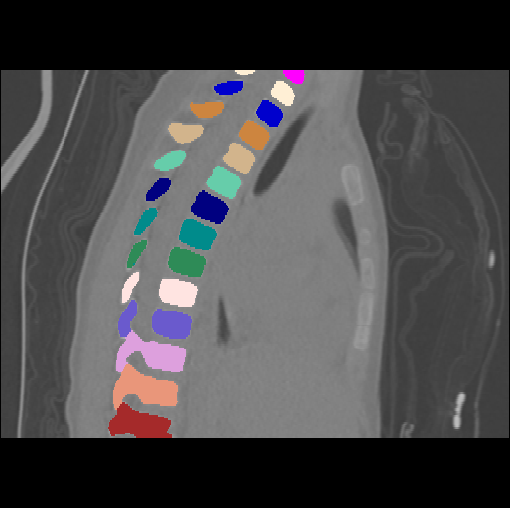}
    \includegraphics[width=\textwidth]{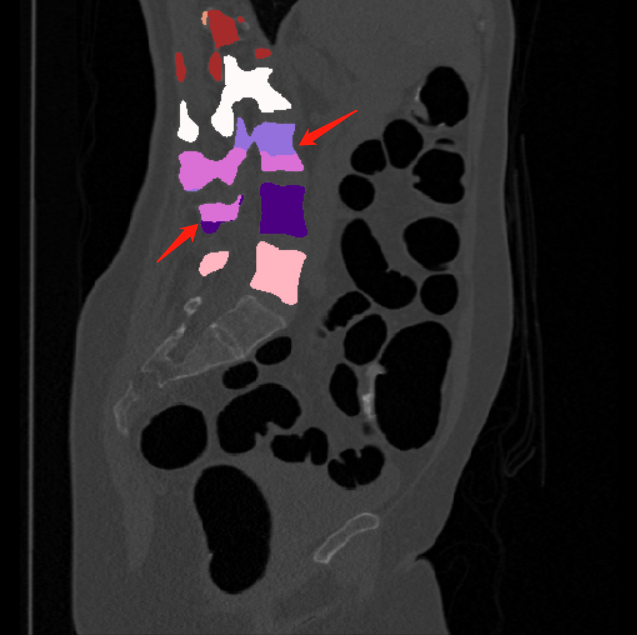}
    Predictions
    \end{minipage}

  \caption{The visualization results on different sub-dataset. The first row and the fourth row, respectively, represent the sub-dataset COLONOG, MSD T10, HNSCC-3DCT-RT, and COVID-19. The last row indicates some failed predictions resulting from spinal diseases: sacral lumbarization and lumbar sacralization. }
  \label{vis-result}
 \end{figure*}

\subsection{Results}
\subsubsection{Quantitative Results.}
We calculate the two metrics of each vertebra, and the results are reported in Table~\ref{tab2}. On the one hand, our experimental results are close to those reported in reference \citep{sekuboyina2020verse} with the same model (nnUnet), verifying the high quality of our annotations. On the other hand, Table~\ref{tab2} shows it is difficult to segment the diseased vertebrae (the DSC of L6 is almost 0). Specifically, the existence of L6 confuses the model, resulting in prediction dislocations (see the last row in Fig.~\ref{vis-result}). Thus, our labeled dataset, which contains many L6 cases, is very valuable for the diseased vertebrae segmentation (we have stated those cases that are hard for annotation in the readme.txt file). Table~\ref{tab2} illustrates that the model trained with our annotations can achieve good performance on our CTSpine1K dataset but a much worse performance on the VerSe Challenge datasets, which explains there is an obvious domain gap between our annotated dataset and the public dataset. We infer the reason is that the COlONOG dataset is based on an empty stomach and colon, confusing the deep learning model by the changes of air content in the abdomen (see Fig.~\ref{difference}). Therefore, our annotations are a good complement to the existing datasets.

\subsubsection{Qualitative results.}
Some visualization results are presented in Fig.~\ref{vis-result}, where we can observe that the baseline model can achieve excellent segmentation results. Nevertheless, some failed predictions occur when spinal diseases exist, especially sacral lumbarization and lumbar sacralization. Besides, the image's resolution of Z direction is closely related to the results, and a lower resolution leads to worse results. Maintaining a reasonable performance for a low resolution is a research challenge. Image superresolution \citep{peng2020saint} might be worth exploring.

\section{Conclusion}
We collect and annotate a large-scale spine CT dataset, including 1,005 CT scans with over 11,000 vertebrae. Furthermore, we validate our dataset using several benchmarking segmentation experiments. We will have more experiments in the future. We believe this work will help stimulate further research on spine-related conditions, including vertebrae segmentation, labeling, and 3D spine reconstruction from biplanar radiographs.

\section*{Author contributions statement}
Y.D., C.W., and S.K.Z wrote the manuscript. The data was annotated by Y.D., C.W., Y.H., Q.L., J.L., S.W.L., M.K.S., Q.Q., S.X.Y., P.B.L, and Y.H.. Y.D. developed the algorithm. H.F.X., C.P.Z., and X.B.W. provided medical guidance to the annotators and reviewed the annotations to ensure their accuracy. All authors reviewed the manuscript. S.K.Z is responsible for the design of the entire work.


\acks{We would like to acknowledge Beijing Jishuitan Hospital for its support in this work.}

%
\ethics{The work follows appropriate ethical standards in conducting research and writing the manuscript, following all applicable laws and regulations regarding treating animals or human subjects.}

\coi{The conflicts of interest have not been entered yet.}

\data{The dataset is publicly available.}



\bibliography{sample}
\end{document}